\documentstyle[epsf]{mn}

\newif\ifAMStwofonts
\def\simlt{\lower.5ex\hbox{$\; \buildrel < \over \sim \;$}}
\def\simgt{\lower.5ex\hbox{$\; \buildrel > \over \sim \;$}}

\ifoldfss
  \ifCUPmtlplainloaded \else
    \NewTextAlphabet{textbfit} {cmbxti10} {}
    \NewTextAlphabet{textbfss} {cmssbx10} {}
    \NewMathAlphabet{mathbfit} {cmbxti10} {} % for math mode
    \NewMathAlphabet{mathbfss} {cmssbx10} {} %  "   "    "
  \fi
  \ifAMStwofonts
    \ifCUPmtlplainloaded \else
      \NewSymbolFont{upmath} {eurm10}
      \NewSymbolFont{AMSa} {msam10}
      \NewMathSymbol{\upi}     {0}{upmath}{19}
      \NewMathSymbol{\umu}     {0}{upmath}{16}
      \NewMathSymbol{\upartial}{0}{upmath}{40}
      \NewMathSymbol{\leqslant}{3}{AMSa}{36}
      \NewMathSymbol{\geqslant}{3}{AMSa}{3E}

    \fi
  \fi
\fi % End of OFSS

\ifnfssone
  \newmathalphabet{\mathit}
  \addtoversion{normal}{\mathit}{cmr}{m}{it}
  \addtoversion{bold}{\mathit}{cmr}{bx}{it}
  \newmathalphabet{\mathbfit} % math mode version of \textbfit{..}
  \addtoversion{normal}{\mathbfit}{cmr}{bx}{it}
  \addtoversion{bold}{\mathbfit}{cmr}{bx}{it}
  \newmathalphabet{\mathbfss} % math mode version of \textbfss{..}
  \addtoversion{normal}{\mathbfss}{cmss}{bx}{n}
  \addtoversion{bold}{\mathbfss}{cmss}{bx}{n}
  \ifAMStwofonts
    \ifCUPmtlplainloaded \else
      \UseAMStwoboldmath
      \makeatletter
      \new@mathgroup\upmath@group
      \define@mathgroup\mv@normal\upmath@group{eur}{m}{n}
      \define@mathgroup\mv@bold\upmath@group{eur}{b}{n}
      \edef\UPM{\hexnumber\upmath@group}
      \new@mathgroup\amsa@group
      \define@mathgroup\mv@normal\amsa@group{msa}{m}{n}
      \define@mathgroup\mv@bold\amsa@group{msa}{m}{n}
      \edef\AMSa{\hexnumber\amsa@group}
      \makeatother
      \mathchardef\upi="0\UPM19
      \mathchardef\umu="0\UPM16
      \mathchardef\upartial="0\UPM40
      \mathchardef\leqslant="3\AMSa36
      \mathchardef\geqslant="3\AMSa3E
    \fi
  \fi
\fi % End of NFSS release 1

\ifnfsstwo

  \newcommand{\s}{\sigma^3}
  \newcommand{\dlnk}{\partial_{{\rm ln} k}}

  \DeclareMathAlphabet{\mathbfit}{OT1}{cmr}{bx}{it}
  \SetMathAlphabet\mathbfit{bold}{OT1}{cmr}{bx}{it}
  \DeclareMathAlphabet{\mathbfss}{OT1}{cmss}{bx}{n}
  \SetMathAlphabet\mathbfss{bold}{OT1}{cmss}{bx}{n}
  \ifAMStwofonts
    \ifCUPmtlplainloaded \else
      \DeclareSymbolFont{UPM}{U}{eur}{m}{n}
      \SetSymbolFont{UPM}{bold}{U}{eur}{b}{n}
      \DeclareSymbolFont{AMSa}{U}{msa}{m}{n}
      \DeclareMathSymbol{\upi}{0}{UPM}{"19}
      \DeclareMathSymbol{\umu}{0}{UPM}{"16}
      \DeclareMathSymbol{\upartial}{0}{UPM}{"40}
      \DeclareMathSymbol{\leqslant}{3}{AMSa}{"36}
      \DeclareMathSymbol{\geqslant}{3}{AMSa}{"3E}
    \fi
  \fi
\fi % End of NFSS release 2

\ifCUPmtlplainloaded \else
  \ifAMStwofonts \else % If no AMS fonts
    \def\upi{\pi}
    \def\umu{\mu}
    \def\upartial{\partial}
  \fi
\fi

\title[Observational constraint on the fourth derivative of the
	inflaton potential]
	{Observational constraint on the fourth derivative of the
	inflaton potential}
\author[Caprini, Hansen \& Kunz]
{Chiara Caprini, Steen H. Hansen \& Martin Kunz\\
University of Oxford, Denys Wilkinson Building, Keble Road, Oxford OX1
	3RH, U.K.}
\date{Draft version \today}
\pagerange{\pageref{firstpage}--\pageref{lastpage}}
\pubyear{2002}

\begin{document}

\maketitle

\label{firstpage}

\begin{abstract}
We consider the flow-equations for the 3 slow-roll parameters $n_S$
(scalar spectral index), $r$ (tensor to scalar ratio), and
$dn_S/d{\rm ln}k$ (running of the spectral index). We show that the
combination of these flow-equations with the observational bounds
from cosmic microwave background and large scale structure
allows one to put a lower bound on the fourth derivative of the
inflationary potential, $M_P^4(V''''/V) > -0.02$.
\end{abstract}

%\begin{keywords}
%\end{keywords}
%\pacs{PACS number(s): 98.80.Cq, 98.70.Vc} 

%%%%%%%%%%%%%%%%%%%%%%%%%%%%%%%%%%%%%%%%%%%%%%%%%%%%%%%%%%%%%%%%%%%%%%

\section{Introduction}

Inflation provides the creation of scalar, vector and tensor
perturbations in the metric; the scalar ones may be the origin of the
formation of large scale structures, the vector ones decayed away, and
the tensor ones gave rise to a stochastic background of gravitational
waves. It is possible to extract informations about the features of
the spectra of these primordial perturbations from observations of the
cosmic microwave background (CMB) anisotropies, or from measurements
of the matter power spectrum from large scale structure (LSS).  The
observable variables are the power spectral indices of the density and
tensor perturbations, $n_S$ and $n_T$, and the overall amplitude of
these, usually denoted by $S$ and $T$. The knowledge of these
quantities allows us first of all to test the inflationary scenario,
by checking if the theoretically predicted consistency relation is
satisfied by the observational values; and then, eventually, to
reconstruct the scalar-field potential. The usual way to express these
spectra is by a Taylor expansion in the deviation from scale
invariance, which can be directly related to the slow roll expansion
in the inflaton potential
\cite{Lidsey:1997np,Martin:2000ak,Leach:2002ar}. Each observable
quantity can then be related to the parameters in the slow roll
expansion.

The evolution of the observables during inflation can be followed by
the flow-equations~\cite{hoffman}, and in~\cite{hoffman} 
it was found, that the lines
$T/S\simeq0$ and $T/S\simeq-5(n_S-1)$ act as attractors for the
evolution in the $(n_S,T/S)$ plane. This result provides a new
relation between the variables $n_S$ and $T/S$, which the authors
suggest can be used as a second consistency relation in interpreting
CMB results. A subsequent analysis found~\cite{steen} that by
combining the observational bounds with the flow-equations, a
non-trivial constraint on the value of the third derivative of the
inflaton potential can be obtained.  Recently Kinney (2002)
generalized the inflationary flow equations to arbitrary order in slow
roll, and numerically integrated them after a truncation at the fifth
order.  In this way one can consider the clustering of 
points in the tridimensional parameter space, including the running of
the spectral index, and hence generalizing the results
of~\cite{hoffman}.

In this paper we will extend the analyses of
refs.~\cite{hoffman,steen,kinney} and show that a combination of the
higher order flow-equations and the observational bound on the
inflationary parameters allows one to put a non-trivial constraint on
the fourth derivative of the inflaton potential.

\section{The flow equations}

Slow roll models are traditionally defined through the three
parameters $\epsilon,\eta$ and $\xi^2$, which are related to the
first, second, and third derivative of the inflaton potential with
respect to the inflaton field $\phi$~(to simplify the expression we
avoid write the reduced Planck mass, $M_P=2.4\times10^{18}$~GeV)
\begin{eqnarray}
\epsilon \equiv \frac{1}{2}\left( \frac{V'}{V} \right)^2 
\, \,  ,  \, \, \, 
\eta \equiv \frac{V''}{V} \, \, , \, \, \,
\xi^2 \equiv \frac{V' V'''}{V^2}  \, .
\label{definitions}
\end{eqnarray}
Assuming that $\epsilon$ and $\eta$ satisfy the flatness conditions,
$\epsilon \ll 1$ and $|\eta|\ll1$, one finds that the scalar spectral
index $n_S$, the tensor to scalar ratio $r$, and the running of the
spectral index $\dlnk\equiv\frac{dn_S}{d{\rm ln}k}$, all observable quantities, can be expressed in
terms of the slow roll parameters as
\begin{eqnarray}
n_S-1 &=& 2\eta-6\epsilon+{\it O}(\xi^2) \, , \nonumber \\ n_T &=& -{r
\over \kappa}\,=\,-2\epsilon+{\it O}(\xi^2) \, ,
\label{epslignt} \\
\dlnk &=& -2\xi^2-24\epsilon^2+16\epsilon\eta+{\it O}(\s)  \, , \nonumber
\end{eqnarray}
with the definition
\begin{eqnarray}
\s&=&{V'^2V'''' \over V^3} \, , \nonumber 
\end{eqnarray}
and where the factor $\kappa$ depends on
the given cosmology~\cite{knox,Turner:1996ge}, in particular on the
value of $\Omega_\Lambda$ and $\Omega_M$. In this paper we will use
the value $\kappa = 5$, corresponding to $\Omega_\Lambda=0.65$ and
$\Omega_M = 0.35$.

The first two expressions in eq.~(\ref{epslignt}) are truncated at
order $\xi^2$, ignoring errors that are quadratic in $\epsilon$ and
$\eta$, and requiring that $|\xi^2|\ll \rm{max}(\epsilon,|\eta|)$
\cite{Lyth:1999xn}. The third expression shows us that every
time we take the derivative of a slow roll parameter we get a quantity
which is one order higher in slow roll, and we therefore need to
introduce the new parameter
$|\s|\ll\rm{max}(\epsilon^2,\epsilon|\eta|,|\xi^2|)$.

From the set of equations~(\ref{epslignt}) one can derive a system of
differential equations which describes the evolution of $n_S$ and $r$
as the inflaton rolls down its potential, and is only weakly
dependent on the form of the potential itself.  By taking, at first
order, ${d\phi \over d{\rm ln}k}=\sqrt{2\epsilon}$, where $\phi$ is
the inflaton field, one gets \cite{Liddle:1992wi,Kosowsky:1995aa}
\begin{eqnarray}
\frac{d \, n_S}{d \, {\rm ln}k} &=&  4 \frac{r}{\kappa} \left[
\left(n_S-1 \right) + \frac{3}{2} \frac{r}{\kappa}
\right] - 2 \xi^2 \, ,
\label{flow1} \\
\frac{d \, n_T}{d \, {\rm ln}k} &=& \frac{r}{\kappa} \left[
\left(n_S-1 \right) + \frac{r}{\kappa} \right] \, .
\label{flow2}
\end{eqnarray}

In our attempt to find a constraint on the quantity $V''''/V$, we will
make the assumption that either $\s$ or $V''''/V$ is constant (see
below), so we will not need to require a better accuracy than the one
given in the set of equations (\ref{epslignt}), and in the corresponding
first order relation ${d\phi \over d{\rm ln}k}=\sqrt{2\epsilon}$. 
The second derivative of the scalar spectral index takes the form
\begin{eqnarray}
{d^2 n_S \over d\,{\rm ln}k^2} &=& 2\s+2\xi^2\eta-24\xi^2\epsilon-192\epsilon^3
+192 \epsilon^2\eta+ \\ \nonumber 
& & \mbox{}-32\epsilon\eta^2+{\it O}(\tau) \, ,
\end{eqnarray}
with $\tau\propto{d\s \over d{\rm ln}k}$, which in our approximation is
negligible or zero.
By using the definitions in eqs. (\ref{epslignt}), we get the new
system of flow equations
\begin{eqnarray}
{d\,n_S \over d\,N} &=& -\dlnk \label{flow3}\\ 
{d\,r \over d\,N} &=& - {r \over \kappa} \left[(n_S-1)+{r \over
\kappa}\right] \label{flow4}\\ 
{d\,\dlnk \over d\,N} &=& -2\s-{1\over2}\dlnk\left(9{r\over\kappa}
-(n_S-1)\right)+ \label{flow5} \\ 
& &\mbox{}2(n_S-1)^2{r\over\kappa}+15(n_S-1)\left({r\over\kappa}\right)^2
+15\left({r\over\kappa}\right)^3 \, , \nonumber
\end{eqnarray}
where we have used the first order expression 
$d \, {\rm ln} k = -dN$, consistent with our assumptions 
in eqs.~(\ref{epslignt}), 
and $N$ is the number of Hubble times (e-folds) until the end of
inflation. 
The parameter $\s$
is related to the fourth derivative of the potential trough
$\s=rV''''/5V$.  In order to close this set of equations, we will
need to assume that either $\s$ or $V''''/V$ can be treated as a
constant throughout inflation. 
The choice between these two assumptions is arbitrary, but
will affect the behavior of the flow in the parameter
space.

\begin{figure}
\begin{center}
\epsfxsize=8.5cm
\epsffile{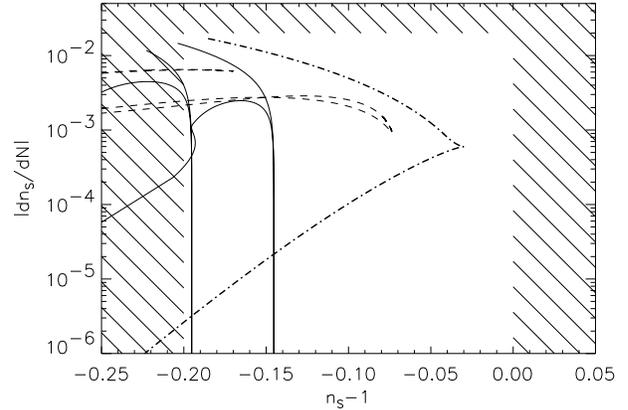}
\end{center}
\caption{The SR validity region flown back 50 e-folds, for different
fixed values of the parameters $\s$ and $V''''/V$. The dot-dashed
line shows the case $\s=0$ (identical to $V''''/V=0$); the two 
dashed lines show, from right to left, the two cases $\s=-10^{-4}$ and
$\s=-3\times10^{-4}$; the solid lines instead represent the case in
which $V''''/V$ is kept constant, and the values are, again from
right to left, $V''''/V=-10^{-2}$ and $V''''/V=-1.8\times10^{-2}$.
The hatched region is excluded by observations.  
It
is clear from the figure, that a slightly larger negative value of the
parameter $\s$ (or $V''''/V$) will not be acceptable because it
doesn't satisfy the observational constraint.}
\label{fig1}
\end{figure}

The observational
bounds which we will use are the ones obtained in \cite{Hannestad:2001nu}
(for similar results see~\cite{al,Hannestad:2001tj,wang,Leach:2002dw})
\begin{eqnarray}
0.8 < & n_s & < 1.0 \, , \label{obsns} \\ 0 < & r & < 0.3 \, ,
 \label{obsr} \\ -0.05 < & \partial_{{\rm ln} k} & < 0.02 \, .
\label{obsdlnk}
\end{eqnarray}   
Such observations can be inverted to give constraints on the
inflationary potential and its derivatives.
The COBE
observations~\cite{Bunn:1996py} gave the first constraint on the first
derivative of the inflaton potential,
\begin{equation}
\frac{V^{3/2}}{M^3 V'} \approx 5 \times 10^{-4} \, ,
\end{equation}
and the bounds above, eqs.~(\ref{obsns}-\ref{obsdlnk}),
provide constraints on the first and second
derivatives of the potential: $|V'/V|<0.25$, $|V''/V|<0.1$.
Hansen \& Kunz (2002) found $V'''/V>-0.2$.

We will study the flow in the three dimensional parameter space $(n_S,
r, \dlnk)$, and by also making use of the observational constraints
(\ref{obsns}-\ref{obsdlnk}), we will be able to induce a limit on the
fourth derivative of the potential, $V''''/V$.  
%We therefore first
%need to derive a flow equation for the variable $\dlnk$, which simply
%is obtained by deriving the third equation of the system
%(\ref{epslignt}) with respect to $d{\rm ln}k$.

\section{Discussion}

In slow roll inflation, the scales relevant for structure formation
crossed outside the horizon roughly $50$ e-folds before the end of
inflation \cite{kolbturner}, and it is at this time that  the
experimental bounds on the observable parameters in eqs. 
(\ref{obsns}-\ref{obsdlnk}) apply. We are going to consider only the case 
of single field inflation, which will end when the slow roll
conditions $V'/V<\sqrt6$ and $V''/V<3$ are
violated. In terms of the value of the parameters $n_S$ and $r$, these
conditions translate into the SR ``validity-region'' \cite{kolbturner,hoffman}
\begin{equation}
{r \over \kappa} < 6 \, \, \, \, \, \mbox{or} \, \, \, \, \, 
\left| \left( n_S-1 \right) + 3\frac{r}{\kappa} \right| < 6 \, .
\label{boundary}
\end{equation}    
Using the evolution equations (\ref{flow3}-\ref{flow5}), we let all
the points of the boundary (\ref{boundary}) flow back 50 e-folds
(we take the value $\kappa=5$).
This is done for different fixed values of the parameter $\s$ (or
$V''''/V$). We can then induce a constraint on this latter by
demanding that, at $N=50$, at least some of the points of the boundary
land inside that region in the $(n_S, r, \dlnk)$ space which is
allowed by observations, eqs.~(\ref{obsns}-\ref{obsdlnk}).  If no
points land inside the space allowed by observations, then we can
deduce that we took an unrealistic value for the parameter $\s$ or
$V''''/V$.

\begin{figure}
\begin{center}
\epsfxsize=8.5cm
\epsffile{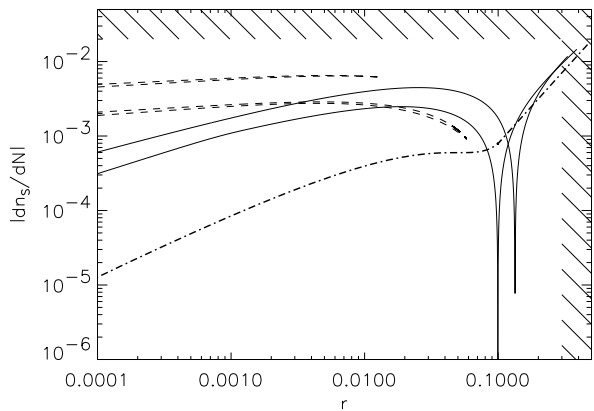}
\end{center}
\caption{Same as fig.~1, but for $r$ and $dn_S/dN$.}
\label{fig2}
\end{figure}

To follow the flow backwards in time, the slow roll conditions
(\ref{boundary}) provide us with the initial values for the two
variables $n_S$ and $r$. As initial condition for the variable $\dlnk$
we will take eq.~(\ref{flow1}), with the extra assumption that
$\xi^2=0$. We verified that the position of the points after 50 e-fold
is virtually independent on the initial conditions chosen, and
therefore this assumption does not affect our results.  Moreover, with
these initial conditions and in the case $\s=0$, all the points of the
boundary flow into the validity region. For other cases (e.g. $\s \neq
0$) we exclude numerically the possibility that a point could first
flow outside this validity region and then later re-enter the region.

In the case where $\s$ is treated as a constant during the last 50
e-folds, we find that for $\s<-3.5\times10^{-4}$ there are no final
points in agreement with observations. This is easily seen in
fig. \ref{fig1}, in which the dashed line which corresponds to the
value $\s=-3\times10^{-4}$ is about to exit from the region given by
the observational limits, eq.~(\ref{obsns}). 
On the other hand, we are not able to induce any limit on the
parameter $\s$ in the case $\s>0$,
because for sufficiently small value of $r$ there will always be
points in the allowed region.

If one instead assumes that $V''''/V$ can be treated as a constant
during inflation, the curve which represent the evolved validity 
region exceeds the
observational boundary if $V''''/V < -0.02$. Also in this case there
will always be acceptable points if $V''''/V$ takes positive values,
however, also in this case only for very small $r$.

\begin{figure}
\begin{center}
\epsfxsize=8.5cm
\epsffile{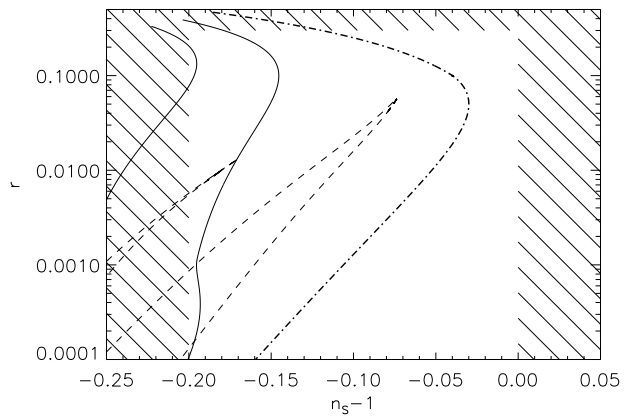}
\end{center}
\caption{Same as fig.~1, but for $r$ and $n_S$.}
\label{fig3}
\end{figure}

In conclusion, we have shown that the combination of the slow roll
equations (\ref{flow3}-\ref{flow5}) and the observational results
(\ref{obsns}-\ref{obsdlnk}) leads to a lower bound $V''''/V>-0.02$,
for single field models which end inflation by breaking the slow roll
conditions (\ref{boundary}). With a larger number of e-folds this
bound becomes slightly stronger.

\section*{Acknowledgements}
We are pleased to thank Pedro Ferreira for useful discussions.  
SHH is supported by a Marie Curie
Fellowship of the European Community under the contract
HPMFCT-2000-00607. MK is supported by a Marie Curie Fellowship of the
Swiss National Science Foundation under the contract 83EU-062445.

\label{lastpage}

\end{document}